\documentstyle[twoside,espcrc2,fleqn,amsmath,epsfig,bm]{article}

\def\gs{{\gamma^*}}
\def\qa{{q\bar q}}
\def\G{{\cal G}}

\def\Fs#1{{}\kern-.45em \not \kern-.12em #1\hspace{.2pt}}

\title{NLO corrections to the photon impact factor\thanks{Talk given
    at the 9th International High-Energy Physics Conference in Quantum
    ChromoDynamics (QCD '02), Montpellier, France, 2-9th July 2002.}}

\author{Stefan Gieseke\address{University of Cambridge, Cavendish
    Laboratory, \\
    Madingley Road, Cambrigde CB3 0HE, United
    Kingdom}\thanks{Supported by the EU TMR-Network `QCD and the Deep
    Structure of Elementary Particles', contract number FMRX-CT98-0194
    (DG 12-MIHT).}}

\begin{document}

\begin{abstract}
  We review the program of the calculation of next-to-leading order
  corrections to the virtual photon impact factor.  Following a brief
  introduction we present some technical aspects for the various
  contributions. Recently obtained results for transversely polarised
  virtual photons are discussed and an outline of how infrared
  divergences are cancelled is given.  Implications of the subtraction
  of leading energy logarithms are discussed.
\end{abstract}

\maketitle

\section{Introduction}
\noindent
Understanding the total cross section for the scattering of two highly
virtual photons, having virtualities $Q_1^2$ and $Q_2^2$ at large
centre-of-mass energy $s$ ($s \gg Q_1^2, Q_2^2$) should be in reach of
perturbative QCD.  The process $\gs\gs\to\rm{hadrons}$ is considered
to be an excellent testing ground for the applicability of
perturbative QCD in the Regge limit \cite{BdRLgsgs,brodsky}.  If the
energy is high enough to validate Regge asymptotics but not too high
in order to suppress unitarity corrections we expect the $\gs\gs$
cross section to be described by the BFKL \cite{BFKL} equation.

To leading logarithmic accuracy (LLA) the predicted cross section,
based on the BFKL equation, rises too quickly with increasing $s$.
The situation is very different at NLO.  The calculation of NLO
corrections to the BFKL Kernel was initiated in \cite{NLOstart} and
finally completed in \cite{FL,CC}.  The corrections were first seen to
be very large.  However, their size is under control when additional
collinear logarithms are taken into account \cite{CCS,salamschool} or
when the kinematical conditions are forced to avoid these extra
logarithms (rapidity vetoes) \cite{schmidt,FRSV}.  The NLO corrections
tend to lower the power rise of cross sections to values that seem to
be compatible with the data \cite{BES,expgsgs}.
\cite{BFKLP1,BFKLP2,BFKLP3,BFKLP4} seem to confirm this although the
$\gs\gs$-cross section is considered using only NLO corrections to the
kernel, ignoring those to the photon impact factor.  These studies,
however, can at best be viewed as an estimate of higher order
corrections since they do not take care of higher order corrections to
the coupling between external particles, virtual photons in the
$\gs\gs$ case, and the NLO BFKL ladder.  In order to make reliable
predictions, being consistent to NLO, the NLO corrections to the
coupling of virtual photons to the exchanged BFKL ladder, described by
the impact factor, has to be taken into account.  These corrections
are currently under study \cite{BGQ,BGK,BCGK} and the status of this
work is reviewed in this contribution.

\begin{figure}[htbp]
  \epsfig{file=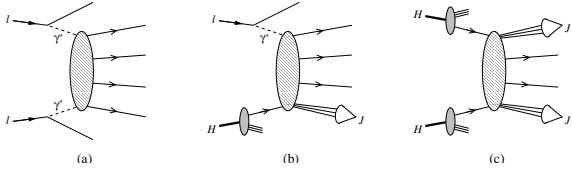,width=\columnwidth}  
  \vspace{-1cm}
  \caption{Hard processes in perturbative QCD at high energy.
  \label{fig:hard}}
\end{figure}

Besides $\gs\gs$ scattering perturbative QCD at high energies may be
studied in any situation where a large rapidity gap between targets is
observed and where a hard scale is involved in observing that final
state (cf.\ Fig.~\ref{fig:hard}).  Prominent examples are the
observation of forward jets in $\gs p$ collisions at HERA
\cite{KMSjets,B_etal_fwjets} or the production of Mueller-Navelet jets
\cite{MNjets} in hadron-hadron collisions.  The coupling of the BFKL
ladder to the relevant jet production vertex at NLO has been finished
recently \cite{BCV1,BCV2,dimitrihere}.

\section{The calculational program}
\noindent
We focus our discussion on the example of $\gs\gs$-scattering which
may serve as the canonical example for a scattering process in
perturbative QCD at very high energy and is sometimes referred to as
the gold-plated process to test BFKL dynamics. 

\begin{figure}[b!]
  \begin{center}    
  \vspace{-1cm}
    \epsfig{file=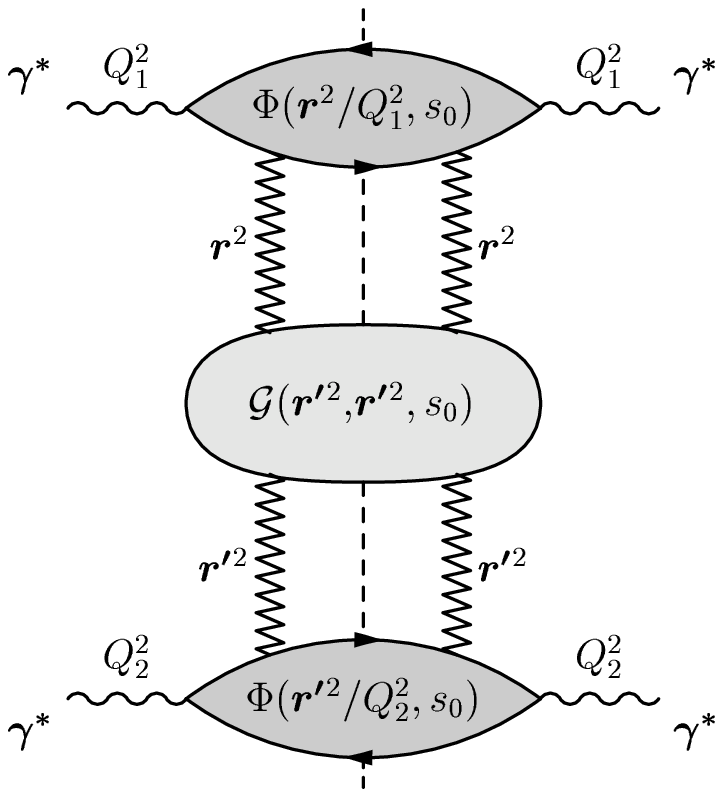,width=5.5cm}
    \vspace{-1cm}
    \caption{Regge factorization of the $\gs\gs$ scattering process.}
    \label{fig:reggefact}
  \end{center}
\end{figure}

\begin{figure}[b!]
  \vspace{-.9cm}
  {\Large
    \begin{displaymath}
      \hspace{-1cm}
      \qquad\Phi_{\gs} = \parbox{4cm}{\epsfig{file=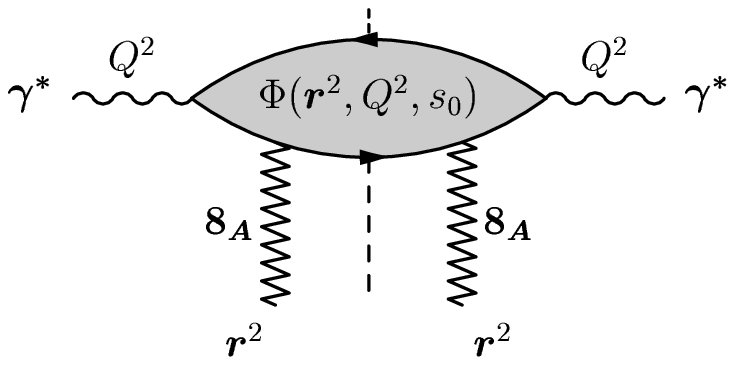,width=4cm}}
    \end{displaymath}
    }
  \vspace{-1cm}
  \caption{The $\gs$ impact factor.}
  \label{fig:gsif}
\end{figure}

At high energy we anticipate Regge factorization and as a result the
total cross section for $\gs\gs$-scattering is written as a
convolution (cf.\ Fig.~\ref{fig:reggefact})
\begin{equation}
  \label{eq:reggefact}
  \hspace{-.8cm}
  \sigma_{\gs\gs}(s) = 
  {\Phi_\gs} \otimes
  {\G_\omega} \otimes
  {\Phi_\gs}\; 
\end{equation}
where $\G_\omega (\bm r^2, {\bm r'}^2, s_0)$ is the Green's function
for the exchange of two reggeized gluons, projected into the colour
singlet state, obtained as a solution of the (NLO) BFKL equation.
$\Phi_\gs$ is the impact factor for virtual photons under discussion.
At leading order, this impact factor (Fig.~\ref{fig:gsif}) is
calculated from cut quark box diagrams: the virtual photon splits into
a $\qa$-pair and the reggeized gluons from the $t$-channel couple to
the $\qa$ pair in all possible ways.

\begin{figure}[t!]
  \begin{center}
    \epsfig{file=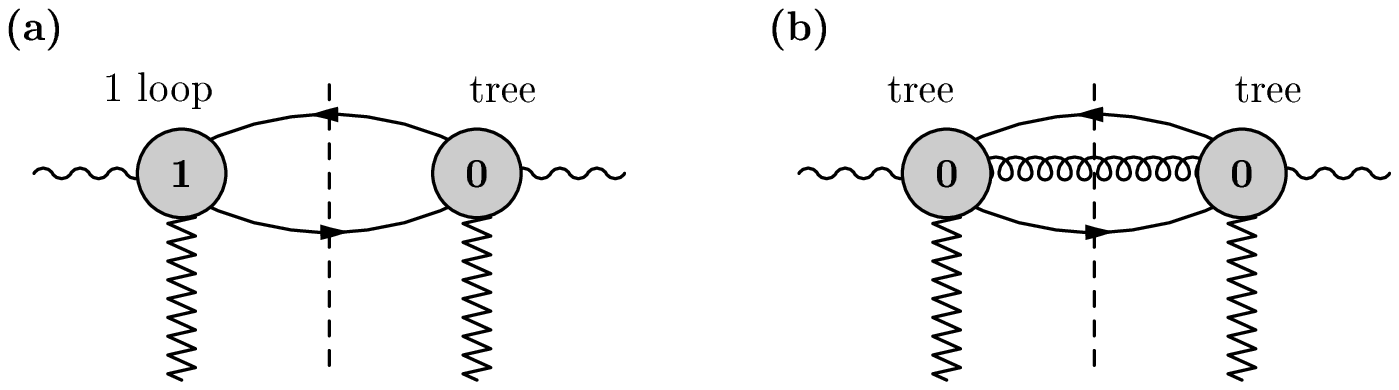,width=\columnwidth}
    \vspace{-1cm}
    \caption{Contributions to the $\gs$-impact factor at NLO. }
    \label{fig:contribif}
  \end{center}
\end{figure}

At NLO we have contributions from virtual corrections to the leading
order diagrams as well as contributions with an additional gluon in
the intermediate state (cf.\ Fig.~\ref{fig:contribif}).  We express
the impact factor in terms of the particle-Reggeon vertices
$\Gamma^{(0)}_{\gs\to\qa}$, $\Gamma^{(1)}_{\gs\to\qa}$ and
$\Gamma^{(0)}_{\gs\to\qa g}$, denoting the coupling of a virtual
photon to Reggeon via a $\qa$-pair at LO, NLO and with an additionally
emitted gluon, respectively.  These vertices result from perturbative
QCD amplitudes with colour octet exchange, taken in the high energy
limit and using Regge factorization.  Expanding the impact factor in
powers of the strong coupling, $\Phi_\gs = g^2 \Phi_\gs^{(0)} + g^4
\Phi_\gs^{(1)}$ we would naively write the NLO impact factor in terms
of these vertices as
\begin{align}   
  \Phi_{\gs}^{(1)} &= \int\!\frac{dM_{\qa}^2}{2\pi} d\phi_{\qa}\,
  2\,{\mathrm{Re}}\,
  \Gamma^{(1)}_{\gs\to\qa}\Gamma^{(0)}_{\gs\to\qa}
\nonumber\\
  &+ \int\!\frac{dM_{\qa g}^2}{2\pi} d\phi_{\qa g}
  \left|\Gamma^{(0)}_{\gs\to\qa g}\right|^2\;.
\end{align}
$dM^2_i$ and $d\phi_i$ denote the invariant mass and the phase space
of the respective intermediate states $i=\qa, \qa g$.

The virtual corrections $\Gamma^{(1)}_{\gs\to\qa}$ have been
calculated in \cite{BGQ}.  We have expressed all loop integrals in
analytic form as an expansion in $\epsilon = (4-D)/2$, quite in
contrast to \cite{FIK} where all integrals are kept as they are.  For
the real corrections, we considered the square of the particle-reggeon
vertex $|\Gamma^{(0)}_{\gs\to\qa g}|^2$, which has been calculated in
\cite{BGK} for longitudinally polarised virtual photons.  In
\cite{BCGK} we recently completed the real corrections by adding the
contributions from transversely polarised photons.  The real
corrections have been considered in \cite{FIK2} as well, but not in a
very suitable form for the task of finally evaluating the impact
factor.

However, calculating the amplitudes as they are does not quite
complete the task.  The individual contributions are still infrared
divergent and have to be combined in order to get the expected
cancellation that has been shown previously \cite{FM}.  At the same
time, one has to consider the subtraction of leading logarithmic
terms.  These are present in the virtual corrections and proportional
to the well-known LLA gluon trajectory function. In the real
contribution to the impact factor they arise as the additional gluon
is emitted with a large rapidity separation to the $\qa$-pair.  Both
of these LLA-terms are individually infrared divergent as well as the
emitted gluon becomes soft.  

In \cite{BCGK} we consider the subtraction of infrared divergences and
leading log terms in combination.  We have extracted the infrared
divergent contributions from real and virtual corrections and defined
suitable subtraction terms.  The difference of our results and the
respective subtraction terms is finite upon integration over the gluon
phase space.  Re-adding the subtracted terms with the integrations
over the gluon phase space performed, explicitly allows us to exhibit
the infrared divergences as poles in $\epsilon$ and cancel them
successfully against those from the virtual corrections.  The final
result for the NLO impact factor reads
\begin{equation*}
  \hspace{-.8cm}
  \Phi_{\gs}^{(1)} = \left.\Phi_{\gs}^{(1,{\mathrm
        v})}\right|^{\mathrm fin} +C_A \left.\Phi_{\gs}^{(1,{\mathrm
        r})}\right|_{C_A}^{\mathrm fin} +C_F
  \left.\Phi_{\gs}^{(1,{\mathrm r})}\right|_{C_F}^{\mathrm fin}
\end{equation*}
\begin{equation*}
  \hspace{-.8cm}
  \quad - \frac{2 \Phi_{\gs}^{(0)}}{(4\pi)^2} \Big\{ \beta_0 \ln\frac{\bm
    r^2}{\mu^2} + C_F\ln(\bm r^2) \Big\}
\end{equation*}
\begin{equation*}
  \hspace{-.8cm}
  \quad +\int\left|\Gamma^{(0)}_{\gs\to\qa}\right|^2   
  \Big\{ C_A \big[\ln^2\alpha(1-\alpha)s_0 -\ln^2 M^2\big]
\end{equation*}
\begin{equation*}
  \hspace{-.8cm}
  \qquad\quad + 2C_F \Big[8 - 3\ln \alpha(1-\alpha) \Lambda^2 + \ln^2
  M^2
\end{equation*}
\begin{equation}
  \hspace{-.8cm}
  \hspace{2.5cm} + \ln^2 \frac{\alpha}{1-\alpha} \Big] \Big\} 
  \frac{dM^2}{2\pi}\frac{d\phi_{\qa}}{(4\pi)^2}\;.
\end{equation}
Therein, $\bm r^2$ is the transverse momentum of the reggeized gluon in
the $t$ channel, $\alpha$ is the light cone momentum fraction of the
outgoing quark and the invariant mass of the $\qa$-pair is denoted
with $M^2$.  In the first line we have the finite remainders from the
calculation from real (r) and virtual (v) corrections. The second line
arises upon renormalization and includes a term proportional to
$\beta_0$, the only term depending on the renormalization scale $\mu$.
The last three lines are the finite remainders of the subtraction
terms.  The dependence on $\Lambda$, characterising the cone in which
the emitted gluon is considered to be collinear, will cancel against a
similar term, implicit in the first line.

The subtraction of the leading logarithmic terms induces a scale $s_0$
which can be translated into a rapidity cutoff beyond which the
emitted gluon will belong to the leading logarithmic term.  However,
since the particular choice of this scale is arbitrary, the NLL impact
factors depend on it.  This dependence was irrelevant at LLA, since a
change in the scale
\begin{equation}
  \hspace{-.8cm}
  \label{eq:log}
  \ln\frac{s}{s_0} = \ln\frac{s}{s_1} + \ln\frac{s_1}{s_0} = 
  \ln\frac{s}{s_1} + \mathrm{NLLA}
\end{equation}
is of higher order w.r.t. the LLA.  These NLLA terms are now taken
care of and may be phenomenologically important.

\section{Outlook and Conclusions}
\noindent
Besides the above discussion of the NLO impact factor, our
calculations have the potential to give further insight into the
photon wave function picture. This picture, in conjunction with the
saturation model has been applied successfully to the description of
both deep-inelastic and diffractive scattering cross sections at HERA,
e.g.\ \cite{GBW1,GBW2}.  First steps in this direction have been done
in \cite{BGK}, showing that an extension of the current picture to a
higher $\qa g$ Fock-state of the virtual photon is in principle
possible.  Further steps in this direction include a consistent
treatment of infrared divergences in configuration space and remain to
be done.

In order to complete the calculation of the impact factor we have to
calculate the phase space integrals over the remaining infrared finite
terms, defined in \cite{BCGK}.  We will express the phase space
integrals in terms of a set of standard integrals.  For first
phenomenological applications this might best be done numerically.

With these results we will be able to calculate the $\gs\gs$ cross
section to NLL accuracy.  In combination with the NLO jet vertex
\cite{BCV1,BCV2,dimitrihere} an interesting variety of
phenomenological applications of the NLO BFKL equation is now
possible.  Numerical results for the cross sections for
$\gs\gs$-scattering, production of forward jets at HERA and
Mueller-Navelet jets at the Tevatron and the LHC are important goals
of future work and remain to be done.

\end{document}